\documentstyle[aasms4,amstex,amsfonts,epsfig,rotating,float,12pt]{article}

\def\refindent{\par\noindent\hangindent=3pc\hangafter=1 }
\def\aa#1#2#3{\refindent#1, A\&A, #2, #3}

\def\apj#1#2#3{\refindent#1, {\it ApJ}, {\bf#2}, #3.}

\def\apjsup#1#2#3{\refindent#1, ApJS, #2, #3}
\def\araa#1#2#3{\refindent#1, ARA\&A, #2, #3}

\def\mnras#1#2#3{\refindent#1, {\it MNRAS}, {\bf#2}, #3.}

\def\msol{{\,M_\odot}}

\def\cm{{\rm\,cm}}

\def\sec{{\rm\,s}}

\def\kelvin{{\rm\,K}}
\def\ergs{{\rm\,ergs}}

\def\gauss{{\rm\,G}}
\def\exp{{\rm\,exp}}
\def\cos{{\rm\,cos}}
\def\sin{{\rm\,sin}}
\def\>{$>$}
\def\<{$<$}

\def\simlt{\lower.5ex\hbox{$\; \buildrel < \over \sim \;$}}
\def\simgt{\lower.5ex\hbox{$\; \buildrel > \over \sim \;$}}
\def\sqr#1#2{{\vcenter{\hrule height.#2pt
      \hbox{\vrule width.#2pt height#1pt \kern#1pt
         \vrule width.#2pt}
      \hrule height.#2pt}}}

\begin{document}

\centerline{Submitted to the Astrophysical Journal (Letters)}
\bigskip
\title{The Hard-X-ray to Gamma-ray Spectrum\\
    in the EGRET AGNs}

\author{Marco Fatuzzo$^{\dag}$ and Fulvio Melia$^*$\altaffilmark{1}}
\centerline{$^{\dag}$Physics Department, The University of Arizona, Tucson, AZ 85721}
\centerline{$^*$Physics Department and Steward Observatory, The 
University of Arizona, Tucson, AZ 85721}





\altaffiltext{1}{Presidential Young Investigator.}


\begin{abstract}
EGRET (20 MeV to 30 GeV) on board the {\it Compton} GRO has observed
high-energy emission from about $\sim 40-50$ Active Galactic Nuclei.  Theoretical models
of this emission based on the upscattering of thermal disk
photons by cooling, relativistic electrons can successfully account
for the EGRET observations, but they predict a considerably
greater X-ray flux than that actually observed in a majority
of these sources.  This inconsistency may be an indication that
the particles are energized during the Compton scattering process,
since the X-ray emission is produced by the lowest energy electrons,
whose density may be relatively small due to the acceleration. Such
a situation may arise as a result of resistive field generation
in electromagnetic acceleration schemes, which we here explore.
A key feature of this model is the assumed existence of a
current associated with the azimuthal component $B_\phi$ of the underlying 
magnetic field by a slight imbalance in the energy distributions 
of outwardly moving, relativistic electrons and protons produced 
at the disk surface via shock acceleration.  The generation of an 
electric field (via magnetic field line reconnection) is thus 
required to maintain the current in the presence of a resistivity 
induced by the radiative drag on the relativistic electrons.
We show that the resulting spectrum can exhibit a significant 
deficit of X-rays compared with $\gamma$-rays.  In addition, 
due to the uni-directional flow of the current associated with
$B_\phi$, this model would predict that the electrons are
energized relative to the protons in the outflow only on one
side of the disk.  They should be decelerated on the reverse side.
As such, we would anticipate that any given blazar should have
a $\sim 50\%$ probability of being $\gamma$-bright, which appears
to be consistent with the observed ratio.
\end{abstract}




%

\section{Introduction}
Many active galactic nuclei (AGNs) produce radio jets or strong radio 
core emission whose origin is still not completely understood.  
Because many quasars are also luminous X-ray sources whose strength 
is correlated with the radio core emission (Browne \& Murphy
1987), it is believed that understanding the high-energy emission, 
which probably originates close to the central, supermassive black 
hole, will yield important insights into the formation of these jets
and the radio emission mechanism. During the past few years,  EGRET on the 
Compton GRO (designed to measure photon energies from 20 
MeV to 30 GeV) has detected about $50$ AGNs, of which $\sim 40$ are
high confidence detections (Mukherjee et al. 1997).  
Most of them are quasars, at least six are BL Lac objects, and at 
least 13 are Optically Violent Variables.  The vast majority are radio loud, 
flat spectrum objects ranging from a redshift $z$ of $0.03$ to more 
than $2.0$.  Typical blazar (i.e., AGNs whose dominant observational
characteristics appear to be due to beaming effects) spectra 
suggest the presence of a break
below $\sim 10$ MeV if the $\gamma$-ray component is to connect
smoothly with the observed X-ray spectrum at lower energies.  
This gamma-ray emission is most likely associated with 
a particle acceleration region near the black hole, thereby providing 
tantalizing clues into the physical processes acting in this environment.

A wide variety of particle energizing mechanisms have been proposed
for the jet formation, ranging from radiation 
pressure-driven flows (Abramowicz \& Piran 1980) to electromagnetically-driven 
self-similar jets (Li, Chiueh \& Begelman 1992).  The most enduring models 
are the electromagnetic ones because they alone appear to have the capability
of accelerating particles to the relativistic speeds necessary to
account for the radiative characteristics of superluminal sources.
The retardation of the particles due to the photon Comptonization 
in the intense radiation field of these compact regions
has been recognized ever since it was
discovered that high-velocity jets could be decelerated by
particle-photon collisions. This mechanism
is sometimes used to account for the low terminal Lorentz factors
($\gamma_\infty\simlt 10$) observed in superluminal sources (Phinney 1982).
Subsequently, Melia \& K\"onigl (1989) demonstrated that the Compton
deceleration due to scatterings with accretion disk photons
could transfer most of the jet's mechanical energy into an
observable flux of high-energy radiation, accounting for both the
low values of $\gamma_\infty$ and the relatively flat X-ray
spectra seen in some sources. Indeed, inverse
Compton scattering models involving a cooling population of leptons
have been successful in reproducing many of the EGRET observations.  
However, the incomplete 
cooling of these leptons in the presence of the disk's nonuniform photon field
results in spectral breaks between the $\gamma$-ray and X-ray 
regimes of $\sim 0.5$ (Dermer \& Schlickeiser 1993), a result that does not 
account for the large variety of
spectral breaks seen in the EGRET AGNs (von Montigny et al. 1995a).
While the presence of additional energy-loss mechanisms can increase the spectral break
to $\sim 0.7$ (Dermer, Sturner \& Schlickeiser 1997), it seems that the underlying 
problem with standard cooling
models is the relative overabundance of low-energy electrons in the assumed
power-law distributions.  To this end, recent theoretical efforts invoking
Comptonization have assumed the presence of an electron acceleration 
process to shift all of the leptons to high energy, thereby producing
a low-energy cutoff in the electron distribution. This energizing
is assumed to occur either before injection into the emission region
(e.g., B\"ottcher, Mause \& Schlickeiser 1997; Dermer, Sturner \& Schlickeiser 1997; 
Sikora et al. 1997) or throughout the interaction zone (Marcowith, 
Henri \& Pelletier 1995; Bednarek, Kirk \& Mastichiadis 1996).  
While these models have had success in accounting for both the X-ray 
and $\gamma$-ray data, the physics of how these conditions arise remains
unsettled, particularly with regard to the specifics of the 
acceleration scheme in the latter class of models.

In this paper, we will be addressing two principal issues.  First, we attempt
to develop a physically self-consistent picture for the mechanism that
produces the required particle distribution with a low-energy cutoff.  Second,
we will follow the evolution of the particle properties with distance away 
from the source of injection, and we will determine the effect of this
on the cumulative photon spectrum. The view studied here is that 
low-energy electrons are first energized 
in a thin transition region in order to 
produce the current associated with the presence of a curled magnetic field
and they then continue to be accelerated {\it during} the Comptonization 
process in order to maintain this current.
Unlike previous models of this type, the acceleration scheme is well-specified 
and the electron distribution is determined self-consistently.
The overall process is assumed to occur within an
approximately force-free magnetosphere, as first described
by Blandford \& Znajek (1977).  Following Melia \& K\"onigl (1989), we
assume the presence of a luminous accretion disk (often invoked
to account for the observed ``blue'' bump) whose
radiation scatters with the accelerating particles.  The magnetosphere
thus retains a small resistivity due to the particle-photon
interactions that couple the acceleration of the particles and their
concurrent radiative emission, but the fields remain close to their
force-free intensities. This resistivity induces a component of the
electric field parallel to the underlying magnetic field
that drives the outflow.  The object of our model is to determine
the individual particle dynamics and the
upscattered radiation spectrum self-consistently.  The merit of such
an approach is evident, in that the energized photon distribution
depends critically on the particle energies.  In \S 2, we describe
a scenario with which the magnetospheric current may be produced,
and we discuss how this current would be a source of $\gamma$-ray
emission.  Our calculations and results are presented in \S 3,
where we also discuss their relevance to the observations.

\section{The Particle Dynamics and High-Energy Emission}
For simplicity, the disk is taken to be a geometrically flat, blackbody 
emitter with inner radius $R_{in}$, outer radius $R_{out}$ and a 
radius-dependent surface temperature as given by (Pringle 1981)
\begin{equation}\label{eq1}
T(R) = T_{max}\left\{7\left({49\over 36}{R_{in}\over R}\right)^3\;\left[1-\left
({R_{in}\over R}\right)^{1/2}\right]\right\}^{1/4}\;,
\end{equation}
where the maximum temperature $T_{max}\equiv (3GM_*\dot M/56\pi\sigma_B 
R_{max}^3)^{1/4}$ (where $M_*$ is the black hole mass,  $\dot M$ is 
the rate of mass transfer through the disk and $\sigma_B$ is the 
Stefan-Boltzmann constant) is attained at the (equatorial) radius $R_{max} 
= (49/36) R_{in}$.  The system is naturally described by a cylindrical 
coordinate system with an origin centered at the black hole and with the $\hat z$
direction normal to the disk.
In this first set of simulations, the magnetic field $B$
threading the disk is assumed to be axisymmetric and, as a result 
of disk rotation, to have an azimuthal component $B_\phi$.

The gravitational energy dissipation power is characterized by
$\dot E_{accr}\equiv GM_*\dot M/R_{in}$.  About half of 
this energy is dissipated throughout the disk.  In the picture developed
here, we assume that energy is transferred to a sub-population
of electrons and protons via shock acceleration
at a rate $\eta\dot E_{accr}$ throughout an active region that extends 
from the inner edge of the disk out to a radius $R_a\ll R_{out}$ 
($\eta$ characterizing the efficiency of the particle energizing 
process). 
Using the above expressions for $T_{max}$ and $R_{max}$,
the rate at which the particles gain mechanical energy is 
expressed as
\begin{equation}\label{eq2}
\dot E_{mech} = 4.2\times 10^{41} \ergs\;\sec^{-1}
\left({\eta\over 0.1}\right)\;
\left({T_{max}\over 5\times 10^4\;\kelvin}\right)^4 \;
\left({M_*\over 10^7\;\msol}\right)^{2}\;
\left({R_{in}\over 3 R_g}\right)^2\;,
\end{equation}
where $R_g = 2GM_*/c^2$ is the Schwarzschild radius.

As a result of shock acceleration, the energized protons are expected to
acquire a relativistic energy distribution characterized
by a power-law in the Lorentz factor $\gamma$.  While the electrons are
also expected to acquire a relativistic energy distribution, the details of 
this process are not well understood.  As such, we make the reasonable 
assumption that the energized electrons are also characterized by a power-law
distribution in $\gamma$ but we do not require that the spectral indices
for the two populations be equal.  In principle, a complex
magnetic field structure would impose a complicated angular distribution 
for the charges.  We note, however, that the observation of GeV photons in 
the EGRET spectrum, coupled with high UV/X-ray luminosities and 
rapid variability, drastically constrain the beam ``size'' of the high-energy
emission (e.g., von Montigny et al. 1995a). This in turn limits the
range of possible magnetic field geometries (i.e., $B_\phi\ll B$), and 
suggests that the dynamics of the outflowing particles 
may be approximated by a more or less unidirectional motion.  (The 
resulting $\gamma$-ray emission does, however, depend on the angular 
distribution of the electrons.  This point will be addressed below.)  
For simplicity, we further assume here that $B_r = 0$ and take the 
energized particles to move directly outward along the $\hat z$ 
direction (since $B_\phi\ll B$). For a uniform active region, the 
initial relativistic electron (proton) distribution function is therefore 
$n_{e(p)}(\gamma; z=0, r) = C\;\gamma^{-\alpha_{e(p)}}$ for $1<\gamma
<\gamma_{u;e(p)}$, where $n_{e(p)}(\gamma; z, r) \;d\gamma$ is the number 
density of electrons (protons) 
at ($r,z,\phi$) with a Lorentz factor between $\gamma$ and $\gamma+d\gamma$.  
Note that since $m_p>>m_e$, almost all of the mechanical energy is
carried by the protons for comparable values of spectral indices.
For simplicity, we assume throughout this paper that $\alpha_p = 2.5$.
The results of our calculations will therefore be insensitive to the value of
the upper bound for the proton distribution $\gamma_{u;p}$.  The value of 
the upper bound $\gamma_{u;e}$ for the electron distribution
is determined by balancing the rate of acceleration within the shock region 
with the combined rate of cooling due to synchrotron and 
Compton processes (Begelman, Rudak \& Sikora 1990).
It is important to note that synchrotron cooling dominates over
Compton cooling within the shock region since the magnetic
field is expected to be highly disordered within that region
(thereby resulting in large pitch angles between the local magnetic field
direction and the electron motion) and the magnetic energy density is 
appreciably greater than the photon energy density.
This therefore leads to a maximum Lorentz factor  $\gamma_u = 
1.16\times 10^6\;({B/ 10^4 G})^{-1/2}$.  For simplicity, we 
adopt the fiducial value $\gamma_u = 10^6$ in our simulations. 
In contrast, since the magnetic field becomes ordered away from 
the disk and the electron motion becomes nearly parallel to the field direction, Compton
cooling processes dominate the dynamics of the flow above the disk. 
For a uniform active region, the value of the  
number density of energized particles is found by equating (2) with the
expression for the mechanical energy flux carried by the protons
$\dot E_p\approx \langle\gamma_p\rangle m_p c^2 2\pi R_a^2 c n_{e0}$, where 
$\langle\gamma_p\rangle$ is the average Lorentz factor for the protons,
$m_p$ is the proton rest mass, and the factor of 2 takes into account
both sides of the disk. Assuming that the flow is charge neutral, the
number density of electrons energized by the shocks can therefore be expressed as 
\begin{equation}\label{eq3}
n_{e0} \approx 6.2\times 10^4 \cm^{-3}\; 
\;\left({\langle\gamma_p\rangle\over 3}\right)^{-1}\;\left({\eta\over 0.1}\right)
\;\left({T_{max}\over 5\times 10^4\;\kelvin}\right)^4
\left({R_a\over 10\, R_{in}}\right)^{-2}\;.
\end{equation}
(We note that for the assumed power-law index $\alpha_p = 2.5$ for the
proton distribution, $\langle\gamma_p\rangle = 3$ so long as $\gamma_{u;p}>>1$.)

The shock mechanism which energizes the electrons and protons acts 
independently of the large scale magnetic field structure through which the
outflowing particles then move.  We therefore assume that the particles must
undergo an adjustment in their distribution within a
(presumably narrow) transition layer in order to produce
the current associated with the presence of $B_\phi$. We do not 
attempt to develop a complete account of this process here but
rather postulate that it occurs within a height $z_0$
due to the action of an electric field generated by magnetic 
field line reconnection so as to attain a self-consistent
magneto-dynamic system.  This assumption is reasonable as long
as the particle energy density $u \approx \dot E_{mech}/\pi R_a^2 c$ 
is much smaller than $u_B\equiv B^2/8\pi$, thereby constraining 
the radius of the active region to a value $R_a \gg 
(8\dot E_{mech}/ B^2 c)^{1/2}$.  A comparison between $n_{e0}$ in 
Equation (3) and the critical number density necessary to produce 
the required current
\begin{equation}\label{eq4}
n_c \equiv {J\over e c}\sim {B_\phi\over 4\pi e R_\phi} 
= 0.56\; \cm^{-3}\;\left({B_\phi\over
10^4 \gauss}\right)\;\left({R_\phi\over R_g}\right)^{-1}\;
\left({M_*\over 10^7 \msol}\right)^{-1}
\end{equation}
(where $R_\phi$ is the characteristic radius of curvature of $B_\phi$), 
then suggests that $n_c$ is probably due to a slight imbalance between 
the electron and proton velocity distributions since $n_c\ll n_{e0}$. 
Throughout the transition zone, the electron distribution undergoes
either an upward or downward shift, depending on the direction of $B_\phi$.
Since the electric field shifts all the electrons over a given distance
by the same $\Delta\gamma$ (ignoring pile-up effects at $\gamma = 1$ for 
a downward shift; these electrons do not contribute to the
$\gamma$-ray emission), the electron power-law distribution 
is maintained throughout the transition zone but the value of the 
lower cut-off $\gamma_l$ may be raised. 
As such, we shall assume a power-law for $n_e(\gamma; z_0, r)$ 
at the transition zone boundary and take $\gamma_l$ to be a free 
parameter characterizing the dynamical processes 
occurring during the transition.  The shift in $\gamma_u$ is small enough 
for us to ignore as long as $\gamma_{l;e}
\ll 10^6$.  Furthermore, as long as $\gamma_{l;e}\ll
m_p/m_e$, the corresponding shift in the proton distribution can also be
ignored. To summarize, it appears that both the negative and positive
charges maintain their outward relativistic motion, but on balance
the electrons are either slightly faster or slightly slower than the
protons in order to account for the required current $n_c$ to maintain
$B_\phi$.  As we shall see, these two situations result in significantly
different $\gamma$-ray emissivities, and this difference may therefore
constitute an important observational signature.

In the absence of any resistive forces, the 
electric field would be quenched once the required current is produced by the
time the particles reach $z_0$.  However, cooling processes (primarily
from inverse Compton scatterings with the disk photons) above $z_0$ 
continue to downshift the energy distribution of the electrons and protons 
(Melia \& K\"onigl 1989).   
As such, this ``Compton resistance'' results in the
generation of an electric field to maintain the current throughout the 
entire emission region.  (We assume that the relativistic proton distribution 
does not change due to Compton upscatterings since by comparison with 
the electrons, the proton Compton scattering
cross section, and hence the proton resistivity, is significantly smaller.)
To determine the effect of these interactions on 
the electron distribution, we first calculate the response of a single electron 
at ($z,r,\phi$) moving with Lorentz factor $\gamma$ through the radiation field 
in the $\hat z$ direction.  Primed quantities denote values in the electron rest 
frame, whereas unprimed parameters pertain to the disk frame (hereafter referred 
to as the lab frame).  The radiation field above the disk is characterized by 
the specific photon number density per solid angle $n_{ph}(\varepsilon, T[R]) 
= (2\varepsilon^2/h^3 c^3) (\exp\{\varepsilon/kT[R]\}-1)^{-1}$, where $\varepsilon$ 
is the lab-frame photon energy and $T[R]$ is the radius-dependent temperature. 
The electron moving through this field scatters $dN$ photons to energies between 
$\varepsilon_s$ and $\varepsilon_s + d\varepsilon_s$ and solid angles between 
$\mu_s \phi_s$ and $[\mu_s + d\mu_s][\phi_s + d\phi_s]$ at a rate (per energy 
per solid angle) 
\begin{equation}\label{eq5} 
{dN\over dt\,d\varepsilon_s\,d\mu_s\,d\phi_s} = \int_{ph} d\varepsilon 
\;\int_{disk}\;\left[{\mu\;R \;dR \;d\Phi\over d^2}\right] n_{ph}
(\varepsilon, T[R])\; \left({d\sigma_{KN}\over d\mu_s' d\phi_s' d\varepsilon_s'}
\right) \;{c (1-\beta\mu)\over \gamma(1-\beta\mu_s)}\;, 
\end{equation} 
where $\beta = (1-\gamma^{-2})^{-1/2}$,
$d \equiv (z^2+[r\;\cos(\phi)-R\;\cos(\Phi)]^2 + [r\;\sin(\phi)
-R\;\sin(\Phi)]^2)^{1/2}$ is the distance between the electron and
disk element $R\;dR\;d\Phi$ located at ($R, \Phi$), and $\mu = z/d$ 
is the cosine of the angle 
between the direction of the scattered photons originating from the 
disk at ($R,\Phi$)  and the electron's direction
of motion. (Note that $\mu$ is also the cosine of the angle between the direction of the
photons incident on the electron and the direction normal to the disk.) 
The expression  in square brackets represents the solid angle subtended
by the disk element $R\;dR\;d\Phi$ at ($R,\Phi$) with respect to the electron.
The differential Klein-Nishina cross-section ${d\sigma_{KN}/ d\mu_s' d\phi_s'
d\varepsilon_s'}$ is evaluated in the electron rest frame.
Using the general expressions relating the lab and rest frame energies  ($\varepsilon' = 
\varepsilon\gamma[1-\beta\mu]$) and angles ($\mu' = [\mu-\beta]/[1-\beta\mu]; 
\phi'=\phi$), one finds the relation 
$d\varepsilon_s d\mu_s d\phi_s / d\varepsilon_s' d\mu_s' d\phi_s' = 
\gamma (1-\beta\mu_s)$, thereby allowing Equation (4) to be  easily 
integrated over all scattered photon energies and solid angles to yield 
the single electron scattering rate.  From this follows the $\gamma$-dependent 
mean-free path length for the electron:  
\begin{equation}\label{eq6}
l_{mfp} (\gamma;z,r) \equiv \left[{dN(\gamma; z,r)\over c\;dt}\right]^{-1} =
\left[
\int_{ph}\;d\varepsilon\;\int_{disk}\;\left[{\mu\;R dR d\Phi\over d^2}\right]
\;n_{ph}(\varepsilon, T[R])\;\sigma_{KN}\;(\varepsilon')\;
(1-\beta\mu)\right]^{-1}\;.
\end{equation}

As a result of a single scattering with a photon of energy $\varepsilon$ and 
direction $\mu$, an electron with initial Lorentz factor $\gamma$ attains a 
lower Lorentz factor $\gamma_s$ in accordance with the energy conservation equation
$\gamma_s = \gamma-[\varepsilon\gamma^2(1-\beta\mu)]/[m_ec^2+\varepsilon\gamma(1-
\beta\mu)]$, where for computational ease, we have set $\mu'_s$ and $\bar\mu'$ to their 
average values (i.e., $\mu'_s$ = $\bar\mu'$ = 0).  Thus, scattering that 
occurs within a distance $dz$ reduces the number density of electrons with Lorentz 
factors between $\gamma$ and $\gamma+d\gamma$ by $n_e(\gamma; z,r)d\gamma\; 
dz/l_{mfp} (\gamma; z,r)$.  Since the particle number is conserved, this 
decrease corresponds to an increase in the number density of electrons with 
a Lorentz factor between $\gamma_s$ and $\gamma_s + d\gamma_s$ of 
$n_e(\gamma; z,r) d\gamma_s\;  dz/l_{mfp}(\gamma; z,r)$.  In the absence 
of a restoring force (i.e., if there were no electric field), the average value of the Lorentz 
factor $\langle\gamma\rangle$ would undergo a gradual downward shift.  As mentioned above, 
however, this downward shift in the electron density function must be balanced 
by the presence of an electric field which maintains the average Lorentz 
factor $\langle\gamma\rangle$ of the distribution, in order to sustain
the current associated with $B_\phi$.  Since the electric field
shifts all electrons moving through the same distance by the same $\Delta\gamma$,
we first calculate the change in the electron distribution function as a 
result of scattering along a path element $dz$ and then linearly shift the 
degraded distribution function upward so as to maintain the value of 
$\langle\gamma\rangle$.  Repeating this procedure stepwise moving outward from 
$z_0$ provides a determination of $n_e(\gamma; z,r)$ for all $z$.

With the electron distribution thus specified, the $\gamma$-ray Compton
emissivity can be determined by integrating Equation (4) over 
the entire scattering electron population.  However, while a unidirectional 
motion was assumed above, the sensitivity of the photon production rate to
the angle between the electron direction of motion and the scattered 
photon direction requires a more realistic treatment of the electron 
angular distribution.  We leave the details of this development
to future work.  The fact that a strong correlation is observed between
$\gamma$-ray emissivity and blazar activity suggests that $\gamma$-rays are 
emitted within a relatively narrow cone (see, e.g., von Montigny et al. 1995a).  
We here assume that this
cone has an opening angle $\theta_c \sim 5^o$, and average the overall 
photon production rate over $1\ge\mu_s\ge \cos\, 5^0$.  Thus, the rate at 
which photons are detected by an observer at a distance $D$ (ignoring 
redshift effects) is given by the expression
\begin{equation}\label{eq7}
{dN_{obs}\over dt\,d\varepsilon_s\,dA} = {2\pi\over (1-\cos 5^o) D^2}\; 
\int_{z_0}^\infty \;dz\;\int_{R_{in}}^{R_a}
\;r\;dr\;\int_1^\infty\; d\gamma \;\int_{\cos\,5^o}^1 d\mu_s\;
n_e(\gamma; z, r)\; {dN\over dt\,d\varepsilon_s\,d\mu_s\,d\phi_s}\;,
\end{equation}
where azimuthal symmetry has been invoked to perform the integral over
$\phi$ and average over $\phi_s$ and the relationship between the 
scattered photon solid angle $d\mu_s \;d\phi_s$ and the detector area 
$dA$ (i.e., $dA = D^2 d\mu_s d\phi_s$) has been used.

\section{Results and Discussion}
\vskip 0.1in
\centerline{\sl (a) Particle Distributions and Characteristic Spectra}
\vskip 0.1in
In this section, we present the results of a calculation for a
$10^7\;\msol$ black hole surrounded by a disk with $T_{max}
 = 5\times 10^4$ K and inner and outer radii $R_{in} = 3 R_g$ and 
$R_{out} = 500 R_g$ (where in this case $R_g \approx 3\times 10^{12}$ 
cm).  Although the luminosity associated with this system ($L\approx
2\times 10^{42}\;\ergs\;\sec^{-1}$) corresponds to the weakest observed
AGNs, qualitatively similar results are expected to follow from
the modeling of more powerful sources, though this will require
substantially more extensive computational resources.  
The active ``shock'' region is 
assumed to extend out to a radius $R_a = 10\;R_{in}$, and the distance 
to the source is set to 100 Mpc. The efficiency $\eta$ is taken
to be $10\%$ and the transition zone boundary $z_0$ is placed at 
an arbitrary value of $5 R_g$.  The dependence of our results on
the actual value of $z_0$ is weaker than their dependence on the
other unknowns in the problem, such as $\gamma_{l;e}$, as long
as $z_0\ll R_{out}$.  We have therefore chosen to concentrate our
simulations on a parameter search corresponding to the dominant
variables.  We note, however, that the radiation field from
the disk is anisotropic, and a careful treatment of the transition
layer may produce some differences with the results shown here.

Before discussing these results, let us first analyze the photon spectrum 
produced by a power-law electron distribution with spectral index $\alpha$
injected at $z_0$ that completely cools via inverse-Compton scattering with 
the disk photons (i.e., the well-studied scenario in which there is
no re-energizing, e.g., due to an electric field).  The resulting radiation spectrum is 
characterized by a power-law ($dN_{ph}\propto \varepsilon_s^{-\Gamma}$)
with index $\Gamma = (\alpha+2)/2$ for energies above the
characteristic value $\varepsilon_{ch}\equiv
3kT_{max}$ ($\approx 10^{-5}$ MeV for the parameter space considered here).  
In practice, however, only electrons with $\gamma$ greater than a critical 
value $\gamma_c$ ($\approx 100$ for these conditions) cool appreciably 
(see, e.g., Dermer \& Schlickeiser 1993; von Montigny et al. 1995a; and 
references cited therein).  As such, the distribution function of the 
outwardly moving electrons will evolve until it exhibits a break at 
$\sim\gamma_c$.  The corresponding photon spectrum, in turn, will exhibit
a break at energy $\sim \gamma_c^2 \varepsilon_{ch}$ ($\approx 10^{-1}$ 
MeV), below which the spectral index will be $\Gamma = (\alpha+1)/2$.
This effect produces a broken power-law spectrum with a shift in $\Gamma$ 
between the X-rays and $\gamma$-rays of $\Delta\Gamma \approx 0.5$ (see 
cooling spectra for $\gamma_l=1$ in figures [3a] and [3b]; hereafter,
$\gamma_l$ always refers to the electrons). 

This situation changes considerably in the presence of an electric field 
that acts above the transition zone to maintain the value of $\langle\gamma\rangle$ 
for the electron distribution injected at $z_0$.  In order to offset the 
energy lost by the high-energy electrons (which cool very efficiently 
via their interactions with the radiation field), an electric field 
is established that shifts all of the electrons upward, including the
low-energy part of the distribution.  The electrons
thus ``bunch'' up at $\langle\gamma\rangle$, which in turn produces a bump in the 
photon spectrum at an energy ${\langle\gamma\rangle}^2\varepsilon_{ch}$.  The overall 
spectral shape of the upscattered photons depends on how the value of 
$\langle\gamma\rangle$ compares with $\gamma_c$.  This point is illustrated in 
Figures 1 and 2.  Figure 1a shows the evolution of the electron distribution 
function at ten different values of $z$ for the model parameters $\alpha = 2$ and
$\gamma_l = 1$.   The curves in figure 1a correspond to the values 
$(z-z_0)/R_g = 0.0027, 0.011, 0.030, 0.080, 0.21, 0.53, 1.4, 3.5, 9.1,$ 
and $23$, with the lowest value corresponding to the power-law curve and 
the subsequent values pertaining to the curves evolving away from this
initial distribution.  The value of $\langle\gamma\rangle$ for this evolution is
13.8.  As expected, the low-energy electrons are shifted upward and 
bunch near $\langle\gamma\rangle$ while the high-energy electrons cool to $\gamma_c$.  
The resulting photon spectrum is shown in Figure 2.  Since $\langle\gamma\rangle 
< \gamma_c$, the  bump at ${\langle\gamma\rangle}^2\;\varepsilon_{ch}$
($\approx 10^{-3}$ MeV) is succeeded by a curve that softens to a power-law 
with spectral index $\Gamma \approx 2$ for energies greater than $\gamma_c^2\varepsilon_{ch}$
(0.1 MeV). The resulting shift in spectral index (i.e., $\approx 0.5$)
between the X-ray and $\gamma$-ray parts of the spectrum is similar to 
that found in the absence of an electric field.

In contrast to the previous case where $\gamma_l = 1$, Figure 1b shows the 
evolution of the electron distribution function for ten different values of 
$z$ when $\alpha = 2$ and $\gamma_l = 100$.   The curves in figure 1b 
correspond to the values $(z-z_0)/R_g = 0.0033, 0.013, 0.033, 0.090,
0.23, 0.58, 1.5, 3.7, 9.3,$ and $23$, with the lowest value corresponding 
to the power-law curve and the subsequent values pertaining to the curves evolving away from this
initial distribution.  In this case, $\langle\gamma\rangle\approx 920$.  The greater value of 
$\langle\gamma\rangle$ over $\gamma_c$ results in all electrons bunching near $\langle\gamma\rangle$.
The resulting photon spectrum is shown in Figure 2.  Since $\langle\gamma\rangle 
> \gamma_c$, the bump at ${\langle\gamma\rangle}^2\;\varepsilon_{ch}$ ($\approx 10$ MeV) 
is succeeded by a power-law with spectral index $\Gamma \approx 2$. This 
spectrum is not well represented by a broken power-law. We note here  that the 
energized distribution ($\gamma_l = 100$) is injected with 
approximately 70 times the energy of the $\gamma_l=1$ distribution
and as such produces $\approx 70$ times
more high energy photons. This point will be elaborated upon
later in the discussion.

\vskip 0.1in
\centerline{\sl (b) Comparison with the Data}
\vskip 0.1in

The high energy photon spectra (30 MeV to 30 GeV) of AGNs observed
by EGRET are well-represented by a power-law with a spectral index 
ranging from 1.4 to 3.0.  This spectrum is a natural consequence of
the inverse Compton scattering of thermal disk photons by a cooling 
power-law distribution of relativistic electrons, though this would
seem to require a range in the electron distribution index 
$\alpha$ between 0.8 and 4.0  which may be broader than that
produced by shock acceleration.  As we have discussed above, the
problem with this scenario arises from the fact that a purely 
cooling electron distribution with $\gamma_l = 1$
produces a spectrum that is not consistent 
with many AGN X-ray observations in that it predicts a larger 
emission of X-rays than is observed (von Montigny et al. 1995a). 
In contrast, it is clear from Figures 1 and 2 that a
low energy cut-off in the electron energy distribution function,
together with an electric field that maintains the average Lorentz factor of
the outflowing electrons in the presence of inverse-Compton 
scattering with the disk radiation, can significantly reduce the predicted
UV and X-ray spectral components.  To illustrate this point in the context 
of the observations, we present in Figure 3a the flux density spectra 
for the model parameters  $\alpha$ = 2 and $\gamma_l$ = 1, 10, and 100, 
along with the multiwavelength data from PKS 0827+243 
and in Figure 3b the flux density spectra for the model          
parameters $\alpha$ = 3 and $\gamma_l$ = 1, 10, and 100, 
along with the multiwavelength data from PKS 1741-038 (taken from
Figure 5 of von Montigny et al. 1995a). The dotted lines extending
from the three different spectra on each figure are for cooling
populations with corresponding values of $\gamma_l$ at the 
transition zone.  
The different scales were chosen for easy comparison of the data
(represented by the left hand scale) to the model results (represented
by the right hand scale). We have not attempted to optimize the
fit here, so the difference between the observed and calculated 
fluxes is presumably due to an imprecise choice of the black
hole mass $M$, source distance $D$, and choice of $\gamma_l$.  

It is evident from our
results that the presence of an energizing region which shifts the
electron population upward is sufficient to produce a
spectrum devoid of X-rays.  The subsequent acceleration
does, however, produce an observable signature in the form 
of a bump around 10 MeV.  As one can see from Figure 3, the main
effect of this feature on the high-energy component of the
spectrum is a turn-up toward lower $\gamma$-ray energies.
There may in fact already be some evidence of this in several 
of the EGRET AGNs discussed by von Montigny et al. (1995a). 

\vskip 0.1in
\centerline{\sl (c) Gamma-ray Faint Versus Gamma-ray Loud Sources}
\vskip 0.1in

A second important prediction of this model is that because the
current associated with $B_\phi$ is due to an imbalance between
the proton and electron outflow rates, the sign of the charge flux 
should be opposite on either side of the disk.  So the electrons 
are re-energized in the transition zone on one side, but the
Lorentz factor of their counterparts on the reverse side must 
be down-shifted. Since the apparent luminosity is
sensitive to the value of $\gamma_l$, AGNs observed from the energized 
side of the disk are clearly much more likely to have a high
(or observable) $\gamma$-ray flux, compared with those observed
from the backside.  The expectation is that only about half of the 
superluminal radio sources should therefore be observable in 
$\gamma$-rays, with a possible additional correlation between 
the spectral break and distance.  

In their analysis of superluminal and strong flat-spectrum
radio quasars that were not seen by EGRET, von Montigny et al.
(1995b) showed that about 35 sources that would otherwise be
expected to be strong $\gamma$-ray emitters were $\gamma$-quiet,
compared to roughly $40-50$ that were detected (either strongly
or marginally).  The lack of evidence for any definitive
characteristics distinguishing these two groups led them to
suggest two primary reasons for the difference: 1) time
variability, and 2) orientation and beam geometry effects.
The scenario we have developed in this paper would point to
a third reason, namely that the undetected sources are for
the most part $\gamma$-quiet due to the uni-directional current
flow associated with $B_\phi$.  As we have discussed above,
the electrons are somewhat decelerated relative to the protons
on one side of the disk, rendering them ineffective in Compton
upscattering the ambient low-energy radiation.  

However, since the observed radio emission associated with
the superluminal motion is produced many parsecs away from
the central engine, we would not expect in this model to see
a correlation between $\gamma$-faintness and a lack of
longer-wavelength blazar activity.  Since most of the mechanical
energy flux is carried by the protons, the energy budget available
for broadband emission in the more extended outflow is roughly
equal on both sides of the disk.  So the $B_\phi$-induced
current flow may offer an elegant explanation for the 
distinction between $\gamma$-loud and $\gamma$-faint blazars.

Finally, on a related matter, one may argue using the curves
in Figure 3 that the X-rays may also be produced by the
same mechanism that accounts for the $\gamma$-radiation, if the 
X-ray portion of the spectrum rises as indicated by the
theoretical curves in this waveband.  Note that to fit both
the $\gamma$-ray and X-ray data, we would need $\gamma_l\approx 
50$ for PKS 0827+243 and $\gamma_l\approx 100$ for PKS 1741-038.
Unfortunately, the available data are not yet sufficient for us
to distinguish between this situation and one in which the
X-rays constitute a completely separate component.  Future
observations may identify the X-ray spectral index with the
required precision for us resolve this issue.

\vskip 0.1in
\centerline{\bf Acknowledgments}

This work was partially supported by NASA grant NAG 5-3075.
We are very grateful to the anonymous referee for his very
helpful comments.  We also thank E. Marietta for helpful 
discussions during the early stages of this work.

\newpage
\section{Figure Captions}
\noindent
Fig. 1a - Evolution of the electron distribution function
for ten different values of $z$ when $\alpha = 2$ and $\gamma_l = 1$.  
The curves correspond to $(z-z_0)/R_g = 0.0027, 0.011, 
0.030, 0.080, 0.21, 0.53, 1.4, 3.5, 9.1,$ and $23$, with the lowest
value indicating the initial power-law and subsequent 
values pertaining to the curves evolving away from this.
The units on the vertical scale are arbitrary.

\noindent
Fig. 1b - Evolution of the electron distribution function
for ten different values of $z$ when $\alpha = 2$ and $\gamma_l = 100$.  
The curves correspond to $(z-z_0)/R_g = 0.0033, 0.013, 0.033, 0.090,
0.23, 0.58, 1.5, 3.7, 9.3,$ and $23$,  with the lowest
value indicating the initial power-law and subsequent 
values pertaining to the curves evolving away from this.

\noindent
Fig. 2 - The photon spectrum as seen by an observer at a distance
of $100$ Mpc for $\alpha = 2$ and the two indicated values of $\gamma_l$.
These spectra are associated with the particle distributions
depicted in Fig. 1.

\noindent
Fig. 3 - (a) Flux density spectrum as seen by an observer at
a distance of $100$ Mpc for $\alpha = 2$ and the three indicated
values of $\gamma_l$ (right-hand scale).  
Multiwavelength observations of PKS 0827+243
are included for comparison (left hand scale).  The data are
taken from von Montigny et al. (1995a).
(b) The flux density spectrum as seen by an observer at
a distance of $100$ Mpc for $\alpha = 3$ and the three indicated
values of $\gamma_l$ (right-hand scale).  
Multiwavelength observations of PKS 1741-038
are included for comparison (left hand scale). The data are
taken from von Montigny et al. (1995a).

%
%
%

{}

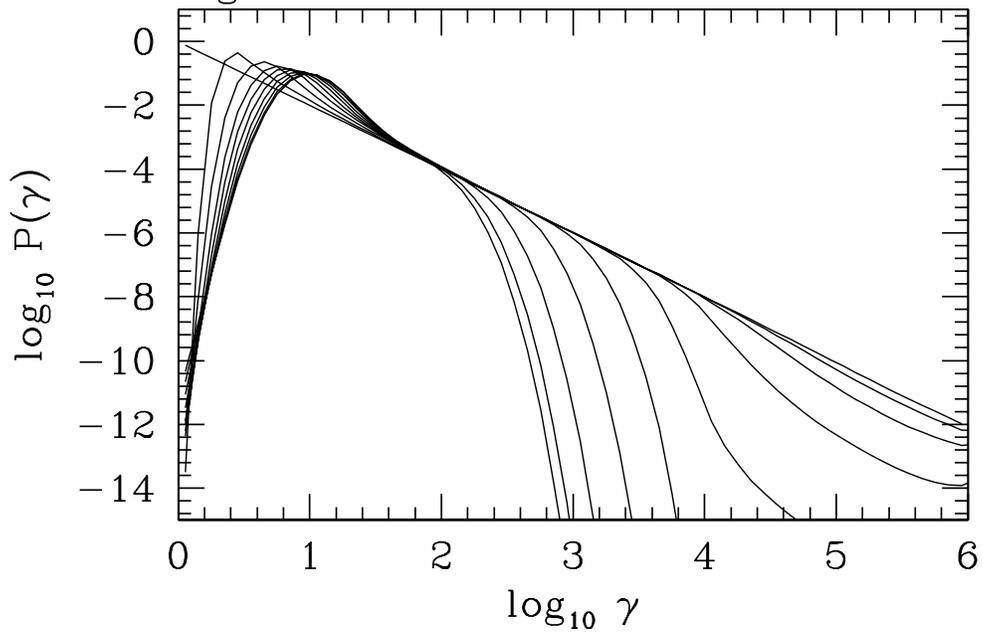
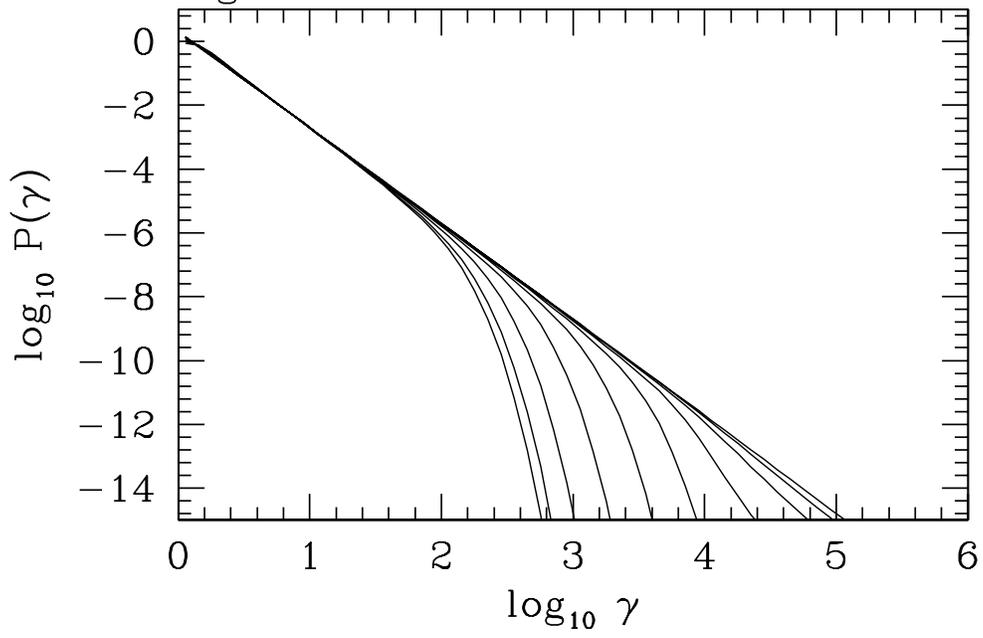
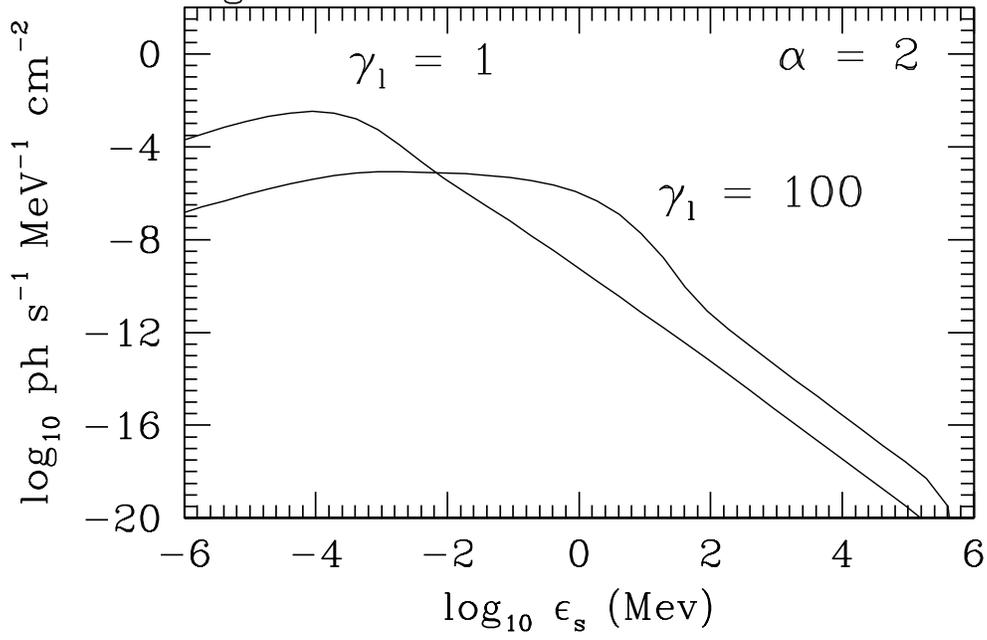
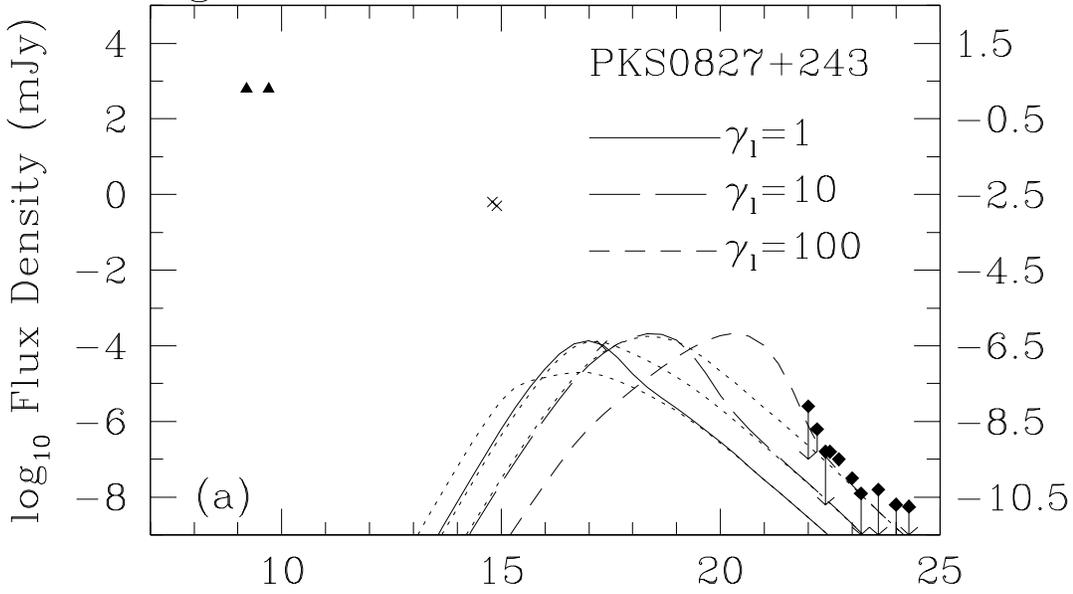
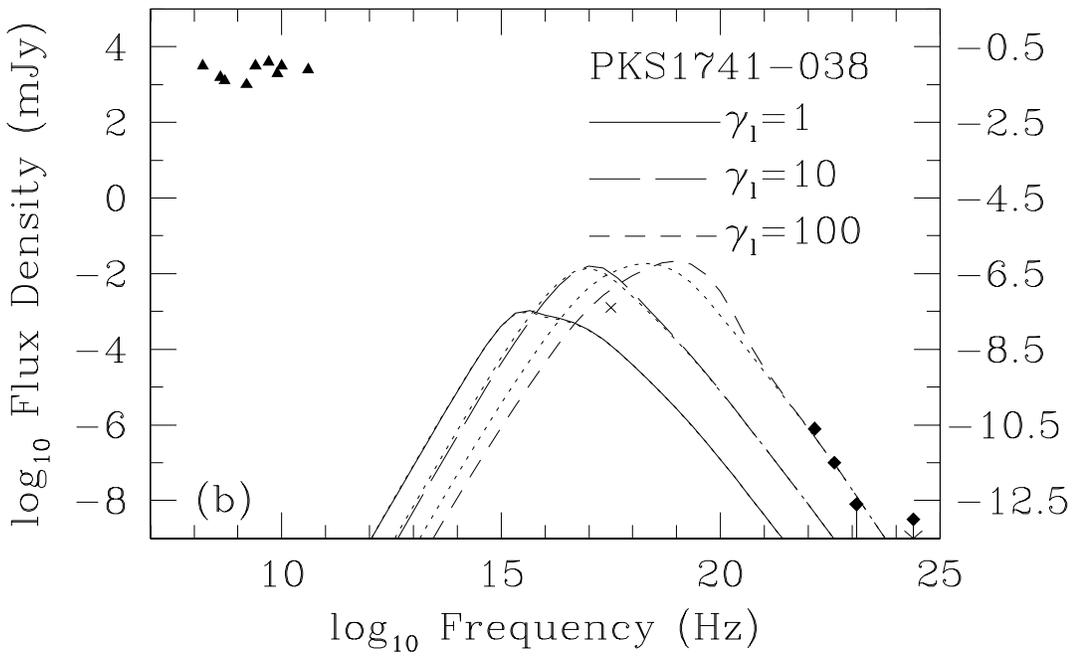
\end{document}